\documentclass{article}

\usepackage{amsfonts}
\usepackage{epsfig}

\begin{document}

\title{\textsc{Discovering Important Nodes
 Through Graph Entropy Encoded
in Urban Space Syntax}}

\vspace{1cm}

\author{ D. Volchenkov and Ph. Blanchard
\vspace{0.5cm}\\
{\it  BiBoS, University Bielefeld, Postfach 100131,}\\
{\it D-33501, Bielefeld, Germany} \\
{\it Phone: +49 (0)521 / 106-2972 } \\
{\it Fax: +49 (0)521 / 106-6455 } \\
{\it E-Mail: VOLCHENK@Physik.Uni-Bielefeld.DE}}
\large

\date{\today}
\maketitle

\begin{abstract}

Potentially influential spaces in the spatial networks of cities
can be detected by means of the entropy participation ratios.
Local (connectivity) and global (centrality) entropies are
considered.  While the connectivity entropy has a tendency to
increase with the city size, the centrality entropy is decreasing
that reflects the global connectedness of cities.
 In urban networks, the local and global properties of nodes
are positively correlated that indicates the
intelligibility of cities.
Correlations between entropy participation ratios
can be  used in purpose of intelligibility measurements and
city networks comparisons.
\end{abstract}

\vspace{0.5cm}

\leftline{\textbf{ PACS codes: 89.75.Fb, 89.75.-k, 89.90.+n} }
 \vspace{0.5cm}

\leftline{\textbf{ Keywords: Complex networks, city space syntax} }

\section{Movement, Intelligibility, and Navigation}
\label{sec:movement}
\noindent

As a city grows
through
 the accumulation of new
buildings and areas,
its street network
emerges which links
 all open spaces
 together and
creating new
 spaces in the
expanding settlement, \cite{SSBeijing}.

Within
the equal socio-economic frameworks
and physical constrains,
a human moves in a direction
that provides him or her the
potential for possible
further movement \cite{Turner}.
J. Gibson
calls such interaction
between humans and environments
 {\it natural
vision} \cite{Gibson}.
The natural vision is a combination of
visual factors affecting
behavior. The next step
has been made in
 by the {\it theory of natural movement}
that
suggests that movement within
a spatial network linking the
buildings will be
determined by the grid configuration
itself \cite{Hillier93}.
A key result of space syntax research is that the pattern
of spatial integration in the urban grid is a key
 determinant of
 pedestrian movement in cities across
  the world,
\cite{Hillier2004}.

This degree of correlation
between aggregate human movement rates
and spatial configuration
is surprising since the
 analysis  of dual city graphs
 incorporate neither
many of the factors considered important in previous models of
human behavior in urban environments (such as the motivations and
the origin-destination information) nor direct account was taken
of the metric properties of space \cite{Penn2001}. Nevertheless,
the robustness of agreement between global integration
(centrality) and rush hour movement rates is now supported by a
number
 of similar studies of
pedestrian movement in different
parts of the world
and in an everyday commercial
work of the {\it Space Syntax Ltd.},
 \cite{Read1997}.
Similar results also exist for vehicular movement
\cite{Penn1998} showing that the spatial
configuration of the
urban grid is in itself a
consistent factor in determining
 movement flows.
 B.Hillier and his colleagues
 \cite{Hillier93}
have shown that the majority of
human-pedestrian movement
occurs along lines of sight,
and that the more {\it integrated}
 (in terms of connection to
other lines of sight) a line is,
the more movement exists along
it. A research has established
that pedestrian movement is more
impacted by the number of turns
than by distance travelled.
Streets from which other
streets can be reached with fewer
direction changes attract
much more people \cite{Implications}.

Land uses which seek movement,
such as markets and retail, then naturally
{\it migrate}
towards higher movement locations,
while others, perhaps residential,
 prefer low movement locations \cite{HillierEconomies}.
 The emerging structure of the spatial
pattern gives rise to a
natural movement pattern,
making some spaces higher
in co-presence than others.
Because more integrated streets
attract more people, they also
tend to attract retail and other
land uses that depend upon the
 volumes of pedestrian traffic,
 and consequently the volumes of
  both pedestrians and
uses are multiplied \cite{Implications}.
Economic growth in the highly integrated streets
feeds back
  onto the structure of
the grid improving its inter-accessibility.
This process will often stabilize at a
certain level related to the original grid
properties that generated the
natural movement in the first place \cite{SSBeijing}.

In the present paper, we investigate the spatial networks (in the
dual representation) of several compact urban patterns
(Sec.~\ref{sec:Spatial_networks}). Potentially influential spaces
can be detected in the urban texture by means of their
contributions into the entropies (by the entropy participation
ratio)
 of the dual  graph
representation of a city (Sec.~\ref{sec:graph_entropy}). The key
observation is that for any graph one can introduce two distinct
classes of entropies. The first class is related to a local
property of nodes in the graph (connectivity). Entropies of
another class are calculated with respect to the different
centrality measures quantifying the global property of nodes in
the graph. Local (connectivity) and global (centrality) entropies
are quite different. For instance, while the connectivity entropy
has a tendency to increase with the city size, the centrality
entropy is decreasing, since a big city usually has "broadways",
the itineraries of prominent centrality
 connecting separated districts of the city (Sec.~\ref{sec:graph_entropy}).

In urban networks, the local and global properties of nodes
are positively correlated that indicates the
intelligibility of cities (Sec.~\ref{sec:Discussion}).
We show that correlations between entropy participation ratios
can be  used in purpose of intelligibility measurements and
city networks comparisons (Sec.~\ref{sec:Discussion}).

\section{Spatial networks of compact urban patterns}
\label{sec:Spatial_networks}
\noindent

A {\it spatial network} of a city is a network of spatial
 elements that constitute the urban environment. They
are derived from maps of {\it open spaces} (streets, places, and roundabouts).
Open spaces may be broken down into components; most simply, these
might be street segments, which can be linked into a network via
their intersections and analyzed as a networks of {\it movement
choices}. The study of spatial configuration is instrumental in
predicting {\it human behavior}, for instance, pedestrian
movements in urban environments \cite{Hillier96}. A set of
theories and techniques for the analysis of spatial configurations
is called {\it space syntax} \cite{Jiang98}.
Although,  in its initial form, space syntax was focused mainly on
patterns of pedestrian movement in cities, later the  various
space syntax measures of urban configuration had been found to be
correlated with the different aspects of social life,
\cite{Ratti2004}.

\begin{figure}[ht]
 \noindent
\begin{center}
\epsfig{file=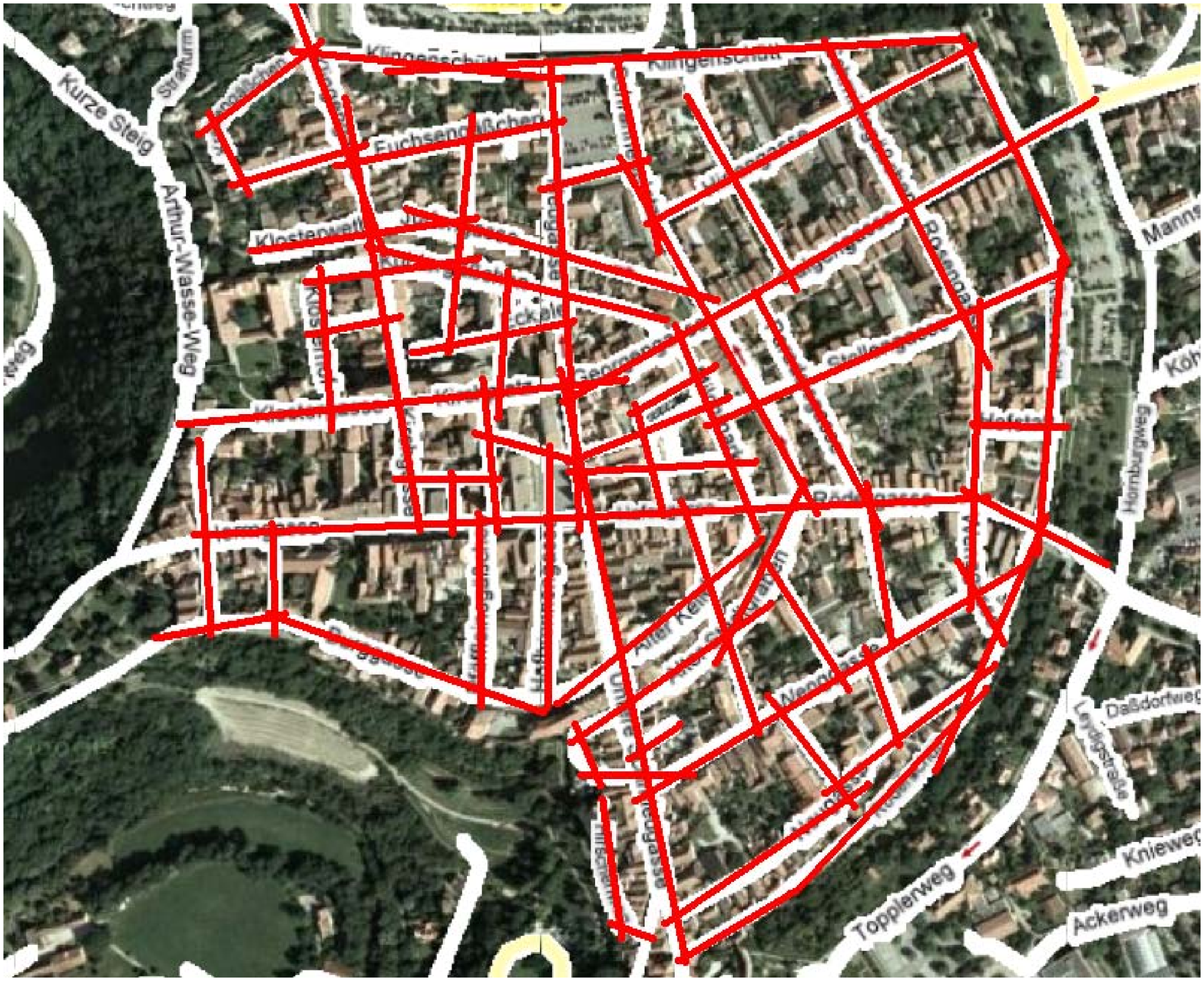, angle=0,width =6.5cm, height =5cm}
  \end{center}
\caption{\small The   axial map drawn for the city of
 Rothenburg ob der Tauber (Bavaria, Germany). }
\label{Fig1_07}
\end{figure}

\begin{figure}[ht]
 \noindent
\begin{center}
\epsfig{file=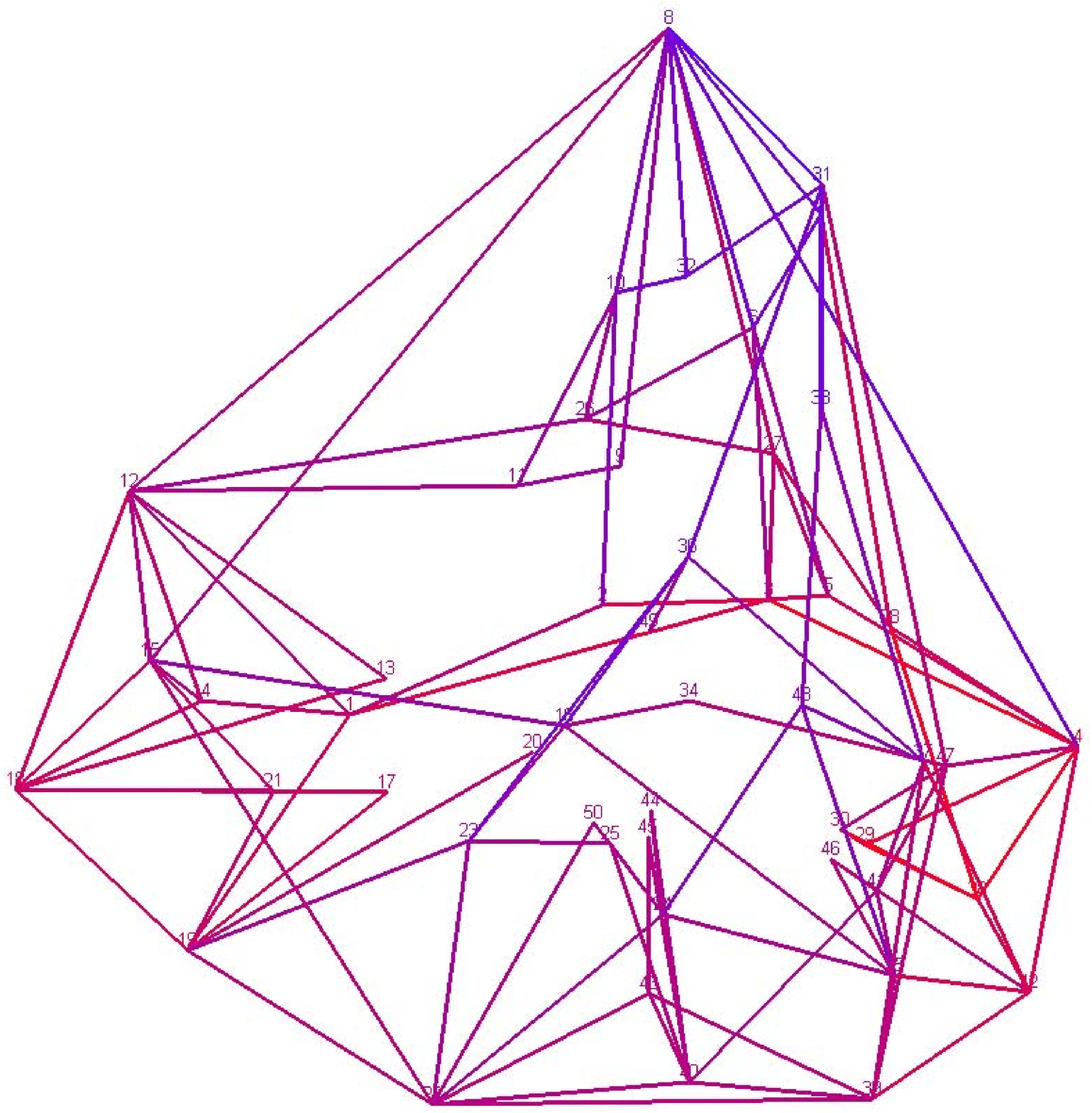, angle=0,width =6cm, height =6cm}
  \end{center}
\caption{\small The morphological representation of the axial map shown
in Fig.~\ref{Fig1_07}. (Rothenburg ob der Tauber, Bavaria, Germany). }
\label{Fig1_08}
\end{figure}

Open spaces are all interconnected, so that one can travel
within open spaces from everywhere to everywhere else.
It is sometime
difficult to decide what should be
 an appropriate spatial element
of the complex space involving large number of
open areas and many interconnected paths.
Decomposition of a space map into a complete set of
intersecting axial lines,  the fewest and
longest lines of sight that pass through every open space comprising any system,
produces an axial map or an overlapping convex map respectively.
In Fig.~\ref{Fig1_07}, we have presented the axial map
drawn for the small city of
 Rothenburg ob der Tauber (Bavaria, Germany).
Axial lines and convex spaces may be treated as the {\it spatial elements}
 (nodes of a morphological graph),
 while either the {\it junctions} of axial lines or the {\it overlaps} of
 convex  spaces may be considered as the edges linking  spatial elements
 into a single  graph unveiling the
topological relationships  between all open elements of the urban space.
In what follows,  we shall call this morphological representation of urban network
as a {\it dual graph}.
 An example of such a morphological representation for the axial map of
 the city of Rothenburg (Fig.~\ref{Fig1_07}) is displayed on Fig.~\ref{Fig1_08}.
We have studied
 five different compact urban patterns.
 Two of them are
situated on islands: Manhattan (with an almost regular grid-like
city plan) and the network of Venice canals
(imprinting the joined
effect of natural, political, and economical factors acting on the
network during many centuries).
In the old city center
of Venice that stretches across 122 small islands in the marshy
 Venetian Lagoon along the Adriatic Sea in northeast Italy,
 the canals serve the function of roads.

We have also considered two organic cities
founded shortly after the Crusades and developed within the medieval
fortresses: Rothenburg ob der Tauber, the medieval Bavarian city
preserving its original structure from the 13$^\mathrm{th}$ century,
and the downtown of
Bielefeld  (Altstadt Bielefeld),
 an economic and cultural center of Eastern Westphalia.

To supplement the study of urban canal networks, we have
investigated that one in the city of Amsterdam. Although
it is not actually isolated from the national canal network, it is
binding to the delta of the Amstel river, forming a dense canal
web exhibiting a high degree of radial symmetry.

The scarcity of
physical space is among the most important factors determining the
structure of compact urban patterns. Some characteristics of studied dual city graphs are
given in Tab.1. There, $N$ is the number of open spaces (streets/canals and places)
 in the urban pattern
(the number of nodes in the dual graphs), $M$ is the number of junctions (the number of
edges in the dual graphs);
the graph {\it diameter},
$\mathrm{diam}(\mathfrak{G})$ is the {\it maximal} depth
(i.e., the graph-theoretical distance)
between two vertices in a dual graph; the intelligibility parameter
(see Sec.~\ref{sec:Discussion})
estimates
navigability of the city, suitability for the passage through it.

\section{Graph entropy encoded
in urban space syntax}
\label{sec:graph_entropy}
\noindent

A major task of
 space syntax
analysis in so far
is the discovery
of potentially influential
spaces in the urban texture
which can attract
a high volume of movement.
It is also important to rank the
commercially potent
spaces in a city
with regard to the rest.

To address such a challenge,
we use an information theoretic model
that combines information theory with
statistical techniques.

The entropy parameter that
is a measure of the uncertainty
associated with a random variable
helps to identify the most
important nodes or a set of such nodes
in a large dual city graph $\mathfrak{G}(V,E)$.

Let $P$ be the probability distribution on the vertex set
$V(\mathfrak{G})$, so that each node $i\,\in\, V$ is
characterized by the probability $p_i\in ]0,1]$ such that
$\sum_{i=1}^N p_i=1$. Within space syntax, the probability
distribution can be defined with relevance to any space syntax
measure discussed in the previous subsections. To be certain, we
consider just two examples of such the distributions.

We define the
 probability
distribution $\pi$ related
to the  local connectivity measure
quantified by
the {\it degree}
of a vertex $v\in V$ is the number
of edges that end at $v$:
\begin{equation}
\label{degree}
\mathrm{deg}(v)= \mathrm{card}\left\{
w\in V: {\ } v\sim w  \right\}.
\end{equation}
The connectivity is a local measures
that shows how well
an open
space is intersecting with other
spaces in the urban pattern,
\begin{equation}
\label{probab_connectivity}
\pi_i=\frac{\deg(i)}{2M},
\end{equation}
in which
$M=|E|$ is the total
number of junctions in the city.
It is worth to mention that
given random walks
defined on the
connected undirected graph
 $\mathfrak{G}$,
such that
 a  walker
moves in one step to
another node randomly chosen
 among all its nearest neighbors,
then
 the probability
distribution
(\ref{probab_connectivity})
 coincides with the unique
stationary distribution of random walkers.
The most important property of the
stationary distribution
 is that
if $\mathfrak{G}$ is not
bipartite, then the distribution
of any node $i$ in random
walks tends to its stationary value
$\pi_i$  as $t\to\infty$.

Another probability distribution
we consider is
\begin{equation}
\label{probab_choice}
p_i\,=\, \frac{\{\# \mathrm{ shortest {\ } paths {\ } through {\ }} i \}}
{\{\# \mathrm{ all {\ }shortest {\ } paths\}}},
\end{equation}
the {\it global choice} measure \cite{glossary}
quantifying the relative structural importance of the node in the
graph. A space $i\,\in\, V$ has a
{\it strong choice} value when many of the shortest
paths, connecting all spaces to all spaces of a
system, passes through it.

Any probability distribution $P$
 leads us directly to the {\it entropy}
(or the structural information content) of a graph:
\begin{equation}
\label{graphentropy}
H(\mathfrak{G},P)=
\sum_{i=1}^Np_i\,\log_2\left( \frac 1 {p_i}\right)
\end{equation}
with the standard convention that $0\,\cdot \log_2 (1/0)=0$. The
information entropy as defined in (\ref{graphentropy}) had been
introduced by C.E. Shannon \cite{Shannon}. We shall refer to the
entropy
 calculated with regard to
the probability distribution
(\ref{probab_connectivity}) as the
{\it connectivity entropy} and the entropy
 calculated
with respect to (\ref{probab_choice})
as the {\it centrality entropy}.

 If all nodes of the graph are
characterized by the equal probability
$p_i$
with respect to the
 probability distribution $P$,
the relevant entropy
should be maximal,
$H_{\max} = N\, \log_2 N$.
For instance,
both entropies calculated
in regard
the probability distributions (\ref{probab_connectivity})
and (\ref{probab_choice}) exhibit  the
maximum value
for
 regular graphs, in which all nodes
are characterized by equal
centrality and connectivity.
Alternatively,
the entropy (\ref{graphentropy}) tends to zero
as all $p_i\to 0$, $i\ne k$,
but one $p_k\to 1.$
The minimal values of both entropies
are achieved for a star graph.
\begin{figure}[ht]
 \noindent
\begin{center}
\epsfig{file=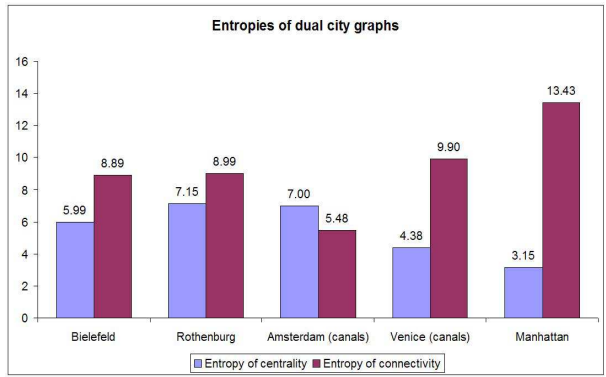, angle=0,width =8cm, height =5cm}
  \end{center}
\caption{\small The centrality and connectivity entropies
of five compact urban patterns.}
 \label{Fig1_15_e}
\end{figure}
In Fig.~\ref{Fig1_15_e}, we have plotted the values of the both
entropies calculated for the five compact urban patterns ordered
with regard to the dual graph sizes. It is then apparent that for
the small organic German medieval cities and for the canal network
of Amsterdam pertaining to a high degree of central symmetry
  the typical values of both
entropies
are proximate.
However,
the
entropies
demonstrate
the alternative tendencies
for much larger urban networks.
The values of
centrality entropy
evidently decrease
for the larger dual city
 graphs
indicating
the presence
of a few nodes
 providing the essentially strong choice.
Alternatively,
 the values of connectivity entropy
blow up for larger networks
bespeaking that the connectivity probability
distributions are smoothed due to the excessive growth of the
number of junctions between streets in large cities.

\begin{figure}[ht]
 \noindent
\begin{center}
\epsfig{file=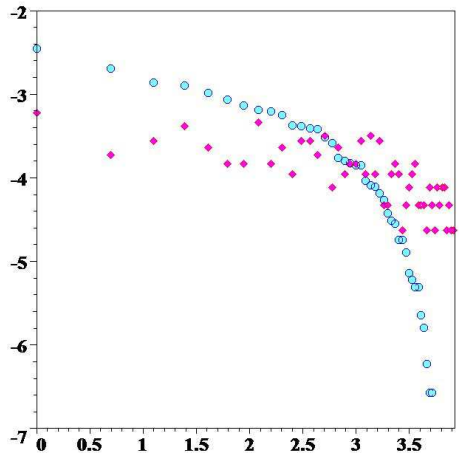, angle=0,width =6cm, height =6cm}
  \end{center}
\caption{\small The rank-EPR  plot (in the log-log scale) of the
dual graph representing the structure of the  Bielefeld downtown.
The solid blue circles are for the EPR regarding the centrality
entropy, and the solid red diamonds are for the connectivity EPR.}
\label{Fig1_16bbb}
\end{figure}

The function
$H(\mathfrak{G},P)$ is continuous with respect to
changing the value of one of the probabilities $p_i$,
symmetric
 being unchanged if open spaces are
re-ordered, and
 additive being
independent
of how the network
is regarded as being
divided into parts.
The last property allows to compute
the entropy of a graph
as a sum of partial entropies pertinent to
 its subgraphs and even
single nodes.
In such a context, the interpretation of important nodes
are those who have the most effect
of the graph entropy.
The importance of a node
in terms of its contribution
into entropy with regard to
the probability distribution $P$
can be estimated by the
 {\it entropy participation ratio} (EPR)
defined as
\begin{equation}
\label{EPR}
\mathfrak{h}_i=\frac{p_i}{H(\mathfrak{G},P)}\,\log_2\left(\frac 1{p_i}\right).
\end{equation}
A key observation, relevant to all
compact urban patterns we have studied
 is that
the prominent nodes
contributing conspicuously into the centrality entropy of the
graph may be inferior with respect to the connectivity entropy
and vice versa.
In order to exemplify
that,
we present in Fig.~\ref{Fig1_16bbb}
the rank-EPR
plot (in the log-log scale) computed for the
dual graph
of the Bielefeld downtown.
Streets have been
sorted according to their centrality
EPR values and then plotted
(solid blue circles).
Their
connectivity EPR
values are given in the same frame
by the solid red diamonds.
Despite the entropy data
in (Fig.~\ref{Fig1_16bbb})
showed considerable variations,
it is obvious that the
both entropies follow a general
tendency that can be
quantified by means of a
correlation coefficient
 indicating
the strength and direction of a
relationship between two data sets. For instance, Pearson's
product-moment linear correlation coefficient  \cite{Pearson}
between the PRE calculated with respect to the centrality and
connectivity entropies for the streets in the downtown of
Bielefeld amounts to $0.7734$. Let us recall that the correlation
is $1$ in the case of an increasing {\it linear} relationship
 between data sets.

\begin{figure}[ht]
 \noindent
\begin{center}
\epsfig{file=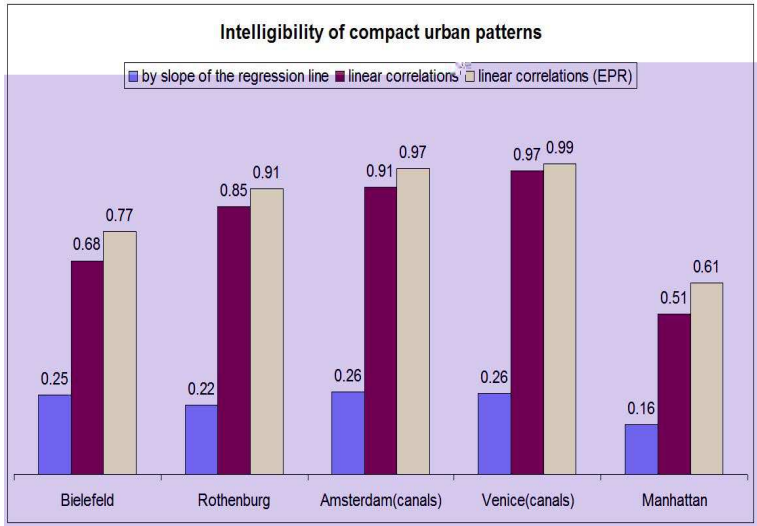, angle=0,width =10cm, height =7cm}
  \end{center}
\caption{\small The comparative diagram of intelligibility indices
calculated for the five compact urban patterns by different
methods.
 }
\label{Fig1_17}
\end{figure}

Proximate values of linear correlation
coefficient can also be found
between the PRE data
for the streets in other
 dual city graphs (see the diagram
in Fig.~\ref{Fig1_17}).
Local and global properties of nodes
come along
 in urban
spatial networks.

The notion that
local and
global
aspects of urban structures
are related is
 at the foundation
 of space syntax.
In particular, it has been suggested in \cite{Penn2001} that if
cities act as {\it mechanisms} for generating contact between
local inhabitants and strangers, then a spatial mechanism at the
basis of this would be likely to include correlations between
local and global movement structures.

\section{Discussion and Conclusion: Intelligibility and Navigation}
\label{sec:Discussion}
\noindent

In space syntax,
 {\it correlations}
between local property
of a space (connectivity)
and global
configurational variables
(integration)
 constitute
a measure of the {\it intelligibility},
the global parameter
quantifying the part-whole
 relationship within the spatial
configuration.
Intelligibility
describes {\it how far} the
depth of a space from the street
 layout as a whole can be inferred
from the number of its direct connections \cite{glossary} that is
most important to way-finding  and perception of environments
\cite{Hillier92,Dalton_Intell}.
 More integrated areas were also
found to be more "legible" by the
residents who perceived their
"neighborhood" to be of a greater size, \cite{Kim99}.

Natural movement relies on
an adequate level of intelligibility
 which has been found
to encourage peoples way-finding abilities.
Spatial integration was found to be
correlated with observed movement, with
the more intelligible area showing
stronger correlations, \cite{Kim99}.

If there is practically no
relationship between how connected
 a node is and how integrated
it is with the overall structure,
 the relationship between space
and movement is weak and the environment seems to be confusing.
While being in a such "unintelligible" layout,  people get lost
more frequently and change directions often that makes navigation
more difficult \cite{RafordHillier}. Movement behavior of humans
in highly complex
 and unintelligible urban areas like
the multi-level urban complexes has been
 investigated in \cite{Chang,ChangPenn}
(the citation appears in \cite{Penn2001}).
The key findings of the research were that
even in highly unintelligible areas movement
was largely predictable from aspects
of the environment. However, whereas in
intelligible urban areas the single
variable of  integration accounted
for the substantial proportion
of variance in movement flows,
in the multi-level complexes a
much wider range of variables
 needed to be taken into account.

In the traditional space syntax approach, the
strong area definition and good intelligibility are
identified in an {\it intelligibility scattergram} and
then by means of the {\it Visibility Graph Analysis}
 (VGA) \cite{HillierEconomies}.
In statistics, a scatter plot is a useful summary of a set of two
variables, usually drawn before working out a linear correlation
coefficient or fitting a regression line. Each node of the dual
graph contributes one point to the scatter plot. The resulting
pattern indicates the type and strength of the relationship
between
 the two variables,
and aids the interpretation
 of the correlation coefficient \cite{Serfling}.

The measure of spatial integration for
each node $i\in\mathfrak{G}$ is usually
taken the
mean distance (called {\it mean depth} in space syntax \cite{glossary}),
 however
the  analysis varies for
specific case studies, and
the precise measures of
the graph are chosen
 to best correlate.

\begin{figure}[ht]
 \noindent
\begin{center}
\epsfig{file=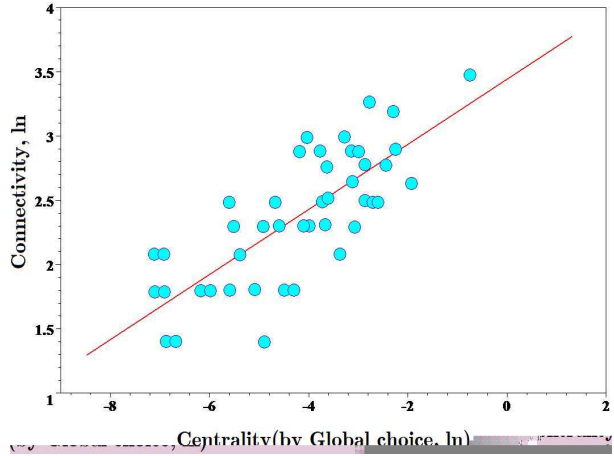, angle=0,width =8cm, height =6cm}
  \end{center}
\caption{\small The intelligibility scatter plot for the Bielefeld
downtown. The slope of the regression line fitting the data by the
method of least squares equals to $0.253$.} \label{Fig1_16}
\end{figure}

We prefer to use the global choice parameter (\ref{probab_choice})
as a measure of spatial integration.
The scatter plot for the
downtown of Bielefeld (in the log-log scale) which shows the
relationship between connectivity and global choice (centrality)
 is
sketched on Fig.~\ref{Fig1_16}.
The pattern of dots (representing the certain open spaces
in the downtown of Bielefeld) slopes from lower left to upper right
that suggests a positive correlation between the connectivity and
centrality variables being studied.
A line of best fit
 computed using the method of linear regression
exhibits the slope $0.253$.
Let us note that the value of Pearson's
coefficient of linear correlations
between the  data samples
of connectivity and global choice
for Bielefeld equals to $0.681$.

We have demonstrated that
the correlations between local and global
properties within the spatial configurations
of urban networks
(intelligibility)
can be quantified by regarding
at least
three different methods.
In the previous subsection, we
have estimated it by means of
  Pearson's
coefficient of linear correlations \cite{Pearson} between the EPR indices
calculated with regard to the centrality and connectivity
entropies. The level of correlations can also be reckoned by the
slope of the regression line fitting the data of the scatter plot
drawn in the logarithmic scale. Eventually, we can directly
compute the correlation coefficient between the uniformly ordered
connectivity and integration values. In order to show the
compatibility of all three methods, we collect the results of all
intelligibility estimations for the five compact urban patterns
that we studied in one diagram (see Fig.~\ref{Fig1_17}).

It is clear from the diagram (\ref{Fig1_17}) that
being an important characteristic related to
a perception of place and navigation within that,
intelligibility can be used in a purpose of
 comparison between
urban networks. The obvious advantage of
 intelligibility is that
it does not depend upon the network size.

The
intelligibility indices estimated by means
of Pearson's
coefficient of linear correlations
between the EPR indices have been given
 in Tab.~1.

\section{Acknowledgment}
\label{Acknowledgment}
\noindent

The work has been supported by the Volkswagen Foundation (Germany)
in the framework of the project: "Network formation rules, random
set graphs and generalized epidemic processes" (Contract no Az.:
I/82 418). The authors acknowledge the multiple fruitful
discussions with the participants of the workshop {\it Madeira
Math Encounters XXXIII}, August 2007, CCM - CENTRO DE CI\^{E}NCIAS
MATEM\'{A}TICAS, Funchal, Madeira (Portugal).

\newpage
\begin{center}
{\bf \small Table 1: Some features of studied dual city graphs}

\vspace{0.3cm}

\begin{tabular}{c|c|c|c|c}
   \hline \hline
 Urban pattern   & $N$ & $M$ & $\mathrm{diam}(\mathfrak{G})$ & Intelligibility
    \\ \hline\hline
 Rothenburg ob d.T. & 50 & 115 &  5& 0.85
 \\
Bielefeld (downtown)& 50 & 142 &  6& 0.68
 \\ 
 Amsterdam (canals) & 57 & 200 & 7& 0.91
\\
 Venice (canals) & 96 & 196 &  5& 0.97
 \\ 
  Manhattan & 355 & 3543 &  5& 0.51
 \\
  \hline \hline
\end{tabular}
\end{center}

\end{document}